# Extracting Urban Sound Information for Residential Areas in Smart Cities Using an End-to-End IoT System

Ee-Leng Tan, Furi Andi Karnapi, Linus Junjia Ng, Kenneth Ooi, Woon-Seng Gan, *Senior Member, IEEE*

*Abstract*— With rapid urbanization comes the increase of community, construction, and transportation noise in residential areas. The conventional approach of solely relying on sound pressure level (SPL) information to decide on the noise environment and to plan out noise control and mitigation strategies is inadequate. This paper presents an end-to-end IoT system that extracts real-time urban sound metadata using edge devices, providing information on the sound type, location and duration, rate of occurrence, loudness, and azimuth of a dominant noise in nine residential areas. The collected metadata on environmental sound is transmitted to and aggregated in a cloud-based platform to produce detailed descriptive analytics and visualization. Our approach in integrating different building blocks, namely, hardware, software, cloud technologies, and signal processing algorithms to form our real-time IoT system is outlined. We demonstrate how some of the sound metadata extracted by our system are used to provide insights into the noise in residential areas. A scalable workflow to collect and prepare audio recordings from nine residential areas to construct our urban sound dataset for training and evaluating a location-agnostic model is discussed. Some practical challenges of managing and maintain a sensor network deployed at numerous locations are also addressed.

## I. Introduction

THE United Nations (UN) estimates that 55% of the world's population lives in urban areas and this figure is projected to reach 68% by 2050 [1]. A growing concern of many modern cities is urban noise, which is usually composed of traffic (road, rail, and air), industrial, construction (on the roads, and buildings), and social (open- and closed-air markets, open- and closed-air eateries, parks, playgrounds, etc.) noise.

Noise pollution is one of the key contributors to loss of environmental quality and lower quality of life. The world health organization (WHO) concludes that diseases related to noise produce a significant loss of healthy life years [2]. Numerous health effects have also been highlighted by WHO and some other studies [2, 3, 4], namely, sleep disturbance, cardiovascular disease, cognitive impairment, and permanent hearing impairment and tinnitus [5]. As noise may change over time and often occurs for only a few minutes or hours, there is a need for continuous monitoring of noise. To aid policymakers to accurately and efficiently access environmental noise occurring in urban areas, a system that monitors and profiles noise by their sound levels and occurrences would be inadequate [6, 7, 8, 9]. Identifying the type of noise can also be used to activate appropriate first responders for timely support and assistance. For instance, law enforcement units can be timely informed of possible riots or illegal gathering of people, and paramedics and ambulances can be dispatched to assist at a car accident.

The internet of things (IoT) is a promising technology that enables many solutions and has been used to address noise pollution in several smart cities. Presently, there is no formal definition of smart city [10, 11, 12], but it is generally agreed that a smart city effectively uses its public resources to provide better services to its citizens. This effort usually leads to the optimization of public services, and management of public resources such as parking [13, 14], lighting [15, 16], and traffic [17, 18]. IoT can also be applied to improve building [19, 20] and waste management [21, 22].

With the rapid advancement of IoT, sensors and embedded processors are becoming smaller, cheaper, and packed with more powerful computational capabilities. Solutions, which were previously difficult to achieve using small form factor and vast deployment, have now been made possible through a combination of IoT and cloud technologies. There have been several attempts to apply such smart devices to monitor and detect sound at the edge. These devices crowdsource sound pressure level (SPL) measurements and generate sound labels across a number of locations over long period of time. One such sensor network designed to address the noise pollution in one of the noisiest cities in North America is the sounds of New York City (SONYC) project [23]. Their acoustic sensor network uses smart sensors constructed using a Raspberry Pi 2 model B and a custom microelectromechanical system (MEMS) microphone module. As of December 2018, a total of 56 sensors were deployed in Greenwich Village, Manhattan, Brooklyn, and Queens. Machine listening was employed to classify the environmental sound at the edge, and their deep neural network (DNN) model was trained and evaluated using their customized UrbanSound8K dataset [24]. This dataset was constructed using recordings obtained from their sensors combined with audio recordings in urban settings downloaded from Freesound [25]. Another project that aimed to develop a real-time system to detect and visualize the acoustic impact of road infrastructure using smart sensors is the (DYNAMAP) project [26]. A two-class sound classifier based on the Gaussian mixture model (GMM) was used to determine the noise contribution from roads and junctions while ignoring noise events unrelated to traffic noise [27]. Another research project referred to as the StadtLärm project [28, 29] focuses on acoustic measurement and sound classification within and around a park located in Jena, Germany. The area under this study is surrounded by two streets that connect the inner city to the nearby highways, trams, and trains. Sounds of activities from nearby restaurants and regular open-air music events were also included in their research. A total of nine sound classes were selected for this study, and their training set was assembled from the dataset UrbanSound8K [24], Tampere University of Technology (TUT) acoustic scenes 2016 database [30], and Freefield 1010 dataset [31].

The availability and serviceability of the sensor network are largely depending on the sensors themselves, which can be increasingly difficult to manage as the sensor network increases in size. A prototype possessing the ability to preempt failures at a rate of 69.1% was developed for the SONYC project [32], but it would be also useful to consider software maintenance of the sensor network to improve system performance using updated algorithms and software modules. Even though azimuths of sound sources are potentially useful in deepening the understanding of the acoustic environment and enhance visualization of the sound sources at the locations of the acoustic sensors, the azimuth of the sound source is absent in many sensor networks. In time-critical scenarios where some sound sources require immediate attention, knowledge of the azimuth of the sound sources would reduce search area and time. The sensor network in the DYNAMAP project was deployed over a large area but focused on road noise. On the other hand, the sensor network of the StadtLärm project was deployed within the vicinity of a park but more sound classes were investigated. It would be interesting to analyze the system performance for multiple sound classes over a large area of deployment. The performance of machine listening is dependent on the annotated dataset used for training, evaluation, and testing. The SONYC and StadtLärm projects leveraged on external sound databases to speed up the annotation process of their dataset.

This paper focuses on addressing acoustic noise problems in urban settings with a special interest in identifying the type of a sound event, as well as its azimuth, SPL, and rate of occurrence. With this rich environmental sound information, policymakers can formulate noise codes that better improve the acoustic comfort in residential areas, especially in a densely populated country such as Singapore. Immediate enforcement action may be taken to react swiftly when certain violation criteria are met, and this decision can be triggered by abnormal sound events and sound levels. Currently, efforts for noise data collection have been addressed through surveys or citizen feedback and complaints to public agencies. However, feedback or complaints are very subjective, which may not be sufficient and are unlikely to be used for formulating regulations and policies.

Therefore, a complaint-driven approach is not effective and is likely to tie up enforcement resources.

In this paper, we present an end-to-end IoT system, which consists of two main domains of edge nodes and a cloud server. A data-driven solution is proposed to tackle acoustic noise problems in residential areas using wireless acoustic sensor nodes (WASN), which infer the type and estimates the azimuth of sound events in the monitored urban environment. These nodes comprise acoustic sensors attached to an embedded processor to perform intelligent sensing around the clock. To tackle the challenges in managing a sensor network deployed over a large area, the sensor nodes are equipped with a self-monitoring mechanism to minimize downtime of the sensor network, and remote updating to automate downloading the latest software to the nodes. To protect the privacy of any human voice captured, only the metadata of the detected sound of interest is transmitted to the cloud server. The training, validation, and testing of sound classification at the edge are carried out with our urban sound dataset solely constructed using the audio recordings collected from the WASN.

Our contributions in this paper can be summarized as follows.
- Developed and deployed an end-to-end IoT system combining machine learning at the edge and cloud platform to analyze urban sound over a large deployment area.
- Combined sound classification and direction-of-arrival (DoA) estimation to detect dominant sound events and estimate azimuths of these sound events at the vicinity of the sensor nodes, and producing a graphical representation of sound events around these nodes.
- Developed a scalable workflow to collect and prepare data samples for training, validating, and testing a location-agnostic model that is deployed at nine residential locations.
- Created an urban sound dataset containing 88,659 labeled sound excerpts of 11 sound classes which are drawn from our taxonomy.
- Incorporated remote updating and self-monitoring features in our sensor nodes to simplify managing and maintaining a sensor network with little or without user intervention.

The rest of the paper is organized as follows. In Section II, the hardware and software architecture of our proposed end-to-end IoT system, cloud management of the WASN, and data visualization are outlined for urban sound sensing. Section III elaborates on the signal processing algorithms used in the WASN, which include the DNN model used to classify the urban sound, and DoA estimation used to estimate the azimuth of the sound. Section IV presents the descriptive analytics of our deployed system, key performance indices of sound classification, and DoA estimation of sound classes at selected deployment sites. Section V highlights some of the key challenges and limitations of our system, together with the conclusions of this work.

## II. ARCHITECTURE OF END-TO-END IOT SYSTEM

In order to build a versatile end-to-end IoT system that provides intelligence and flexibility in extracting key urban sound information from the environment, we propose a system architecture as shown in Fig. 1.

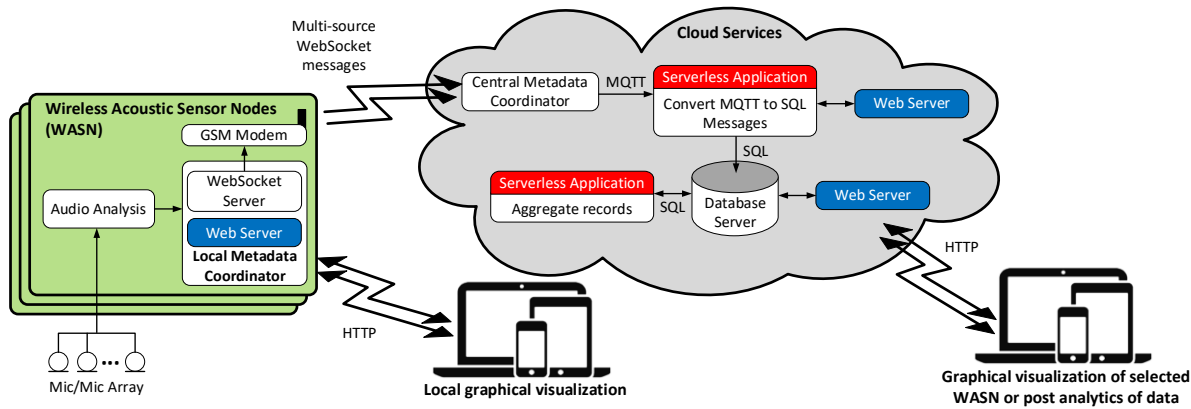

Fig. 1. Overview of end-to-end IoT system architecture for extracting urban sound information.

The environmental sound information is captured by either one or an array of acoustic MEMS sensors (or microphones) and digitized before being passed to a local processing unit that performs edge analytics. On-site sound analysis performed by the WASN includes sound classification, DoA estimation of the detected sound events, computation of SPL, and derives metadata of the detected sound event. A local metadata coordinator in the WASN packs this set of metadata into a JavaScript object notation (JSON) format. Using a GSM modem, the local metadata coordinator wirelessly transmits the metadata in WebSocket messages to the central metadata coordinator in the cloud. The central metadata coordinator relays WebSocket messages from a selected WASN to a web server hosting visualization to gain insights and perspective on the noise characteristics at a particular location. The central metadata coordinator also parses the multi-source WebSocket messages as an array of JSON-formatted metadata. This metadata is transmitted over message queuing telemetry transport (MQTT), a lightweight machine-to-machine connectivity protocol, to a serverless application. This application converts the MQTT messages into structured query language (SQL) commands and stores the metadata into SQL tables. At every 10-minute interval, another serverless application is automatically invoked to consolidate the metadata records by grouping them into snapshot tables according to the hour, day, week, and month. Each consolidated record represents the total number of occurrences of a sound class and average, minimum, and maximum SPL within an hour. The aggregated data can be used to perform a series of descriptive analytics. The metadata records can also be used to generate real-time alert notifications to relevant authorities, or remotely activate alert systems to the sensor location. A web application is developed to provide an interface to visualize the metadata on desktop and mobile devices. This web application is accessible over the cloud through an internet browser on these devices, and hence, it provides a display interface for immediate assessment by the end-user without reliance on a specific platform of implementation. The following sub-sections detail the main hardware, software, and cloud components of the end-to-end IoT system.

### A. Hardware Edge Component for Urban Sound Sensing

The WASN should possess computation, storage, and communication capabilities. In the case of our sensor node, it infers the sound class and estimates the DoA of the dominant sound source in the urban environment at a particular time instance, and communicates with the cloud server to report critical incidents and monitor noise conditions. Therefore, there must be sufficient compute-storage capability of the sensor to perform these complex tasks. WASN can be deployed in common areas where there are high human traffic and activities or in remote areas, where WASN can provide added surveillance for public safety or important facilities.

WASN can be categorized into either mobile or static. The former uses a mobile phone to pick up the acoustic information on the move. Presently, our deployments only use static WASN, where sensor nodes are fixed in a location such as at a lamp post and on a wall. Therefore, it is important to consider the suitability of deployed locations of the static WASN.

There are two hardware configurations for our WASN, which differ in the microphone selection. The first configuration uses a MEMS microphone to perform sound event classification and other sound pressure sensing functions. The second configuration uses a circular planar MEMS microphone array that performs identically to the first configuration but includes DoA estimation of the dominating sound source. The main hardware components of our WASN are:

- Embedded ARM processor [33] running on Raspbian Buster Lite operating system.
- Two configurations of MEMS microphone:
  1. Single-channel unit [34] connected to the embedded processor through I2S connection.
  2. 7-channel array [35] connected to the embedded processor through USB connection.
- GSM modem [36] in two operating modes:
  1. Hilink mode: Provides an ethernet interface for internet connectivity.
  2. Stick mode: Provides a point-to-point interface and maintains continuous internet connectivity. This mode avoids the implementation of the gateway since each sensor node pushes its data directly to cloud services using the attached GSM modem.

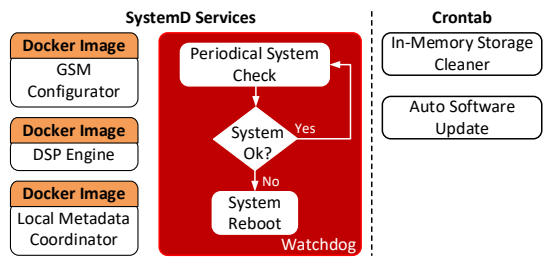

Fig. 2. Block diagram of middleware and software of sensor node.

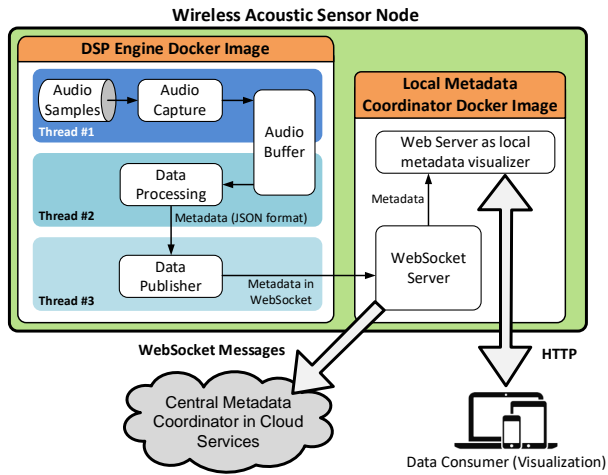

Fig. 3. Key blocks in DSP engine and local metadata coordinator Docker images.

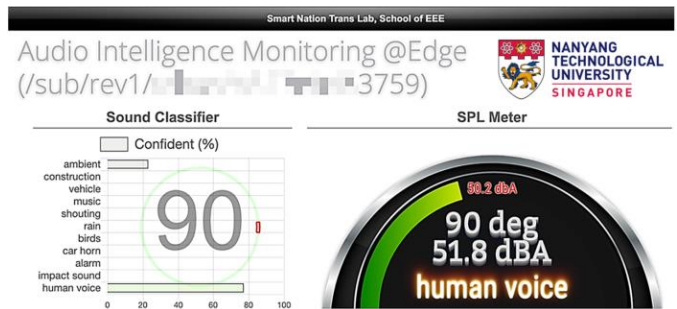

Fig. 4. Data visualization on internet browser showing noise classification, DoA, SPL, and $L_{Aeq}$.

The sensor node is housed inside a splash-resistant enclosure with an IP65 rating and the custom-built microphone housing is attached to the bottom of the enclosure using an IP65 cable gland.

*B. Sensor Node Software*

The software of our WASN is running on Raspbian Buster, a multi-threaded operating system. Low-level maintenance software components, such as watchdog monitoring, memory clean up, and automatic download/upgrade of signal processing software are set to run directly under Raspbian Buster. Ideally, the sensor node should self-recover when it faces any unexpected conditions causing it to stop operating. To achieve this functionality, the built-in hardware watchdog timer of the embedded processor is configured to perform system health checks periodically and to reset the hardware whenever needed.

The software in the WASN can be categorized into two groups, as shown in Fig. 2. The first group of software runs inside SystemD services, which consists of the GSM configurator, digital signal processor (DSP) engine, local metadata coordinator, and watchdog monitoring. The second group of software runs through time-triggered middleware (through Linux Crontab). Triggered by the daily Crontab, each WASN looks for the latest Docker images from the cloud repository and updates itself with the latest Docker image when it is available. Docker images are lightweight, standalone, and executable packages needed to run relevant services provided by the Linux kernel. The software implementation of our WASN, namely, GSM modem configurator, DSP engine (which performs sound event classification, SPL computation, and DoA estimation), and local metadata coordinator with a built-in metadata visualizer is packaged into three Docker images. The daily Crontab also triggers the removal of older files in the software system when the available system memory of the WASN is low.

TABLE I METADATA FROM SENSOR NODE

| Metadata | Description |
| --- | --- |
| *Class index* | Inferred class index (actual class name is referred from look-up-table stored in ClassLUT table in database) |
| *Score* | Confidence score array for sound classification |
| *Average Confidence* | Average confidence score over one sec |
| *SPL* | A-weighted instantaneous SPL |
| *LAEQ* | One-minute equivalent SPL (A-weighted) |
| *Timestamp* | Timestamp of detected sound event |

The DSP engine Docker image is implemented in Python using a Tensorflow/Keras framework [37]. The sound captured by the MEMS microphone is sampled at 44.1kHz and buffered into frames, with each frame having seven channels and each channel contains 8,192 audio samples. Each audio sample is quantized into a 16-bit fixed-point number. The DSP engine utilizes multiple-threading features in Python to split the computational tasks among audio capturing, data processing, and data publishing, as shown in Fig. 3.

Thread #1 captures the data from the microphone audio buffer, prepares the data into a suitable format, and repacks the formatted data into a multi-frame buffer. Thread #2, which is the data processing thread, performs sound classification and computes A-weighted SPL, equivalent sound average ($L_{Aeq}$), azimuth, and timestamp of the detected sound class. The output of the data processing block in Thread #2 is referred to as the metadata (in JSON format) and a summary of the metadata produced by the data publisher is presented in Table I. The metadata is continuously pushed out to the central metadata coordinator by the data publisher block (Thread #3). The data publisher serves as the final output formatter encapsulated inside the DSP engine Docker image. As shown in Fig. 3, the data publisher establishes a connection to a local WebSocket server and then pushes each set of metadata out to the central metadata coordinator using the WebSocket protocol.

Local metadata visualization plays an important role during on-site testing. Local metadata visualization is implemented as a web application and graphically presents the metadata of the node in graphs and charts, as shown in Fig. 4. By directly connecting to the sensor node and without going through the cloud, WebSocket messages can be viewed at the deployment site through an internet browser on a client device. The sensor node itself serves as a WiFi access point to form a private network with the client device.

Fig. 5. Aggregating metadata records into hourly records. Rows in green represent records of the "HumanVoice" class, being processed from raw records (left table) to aggregated records (right table).

TABLE II TAXONOMY OF SOUND CLASSIFIER

| Category | Sound Label | Description |
| --- | --- | --- |
| Machinery | Vehicle | Sounds from engines of vehicles, such as motorcycles, cars, buses, lorries, and aircraft. Also includes engine/motor sounds made by non-mobile machines, such as fruit shredders, pressure cleaner motors (motor pumps), leaf blowers, and screeching brake sounds produced by vehicles. |
| | Construction | Non-engine sounds produced from construction machinery, such as jackhammers, breakers, and powered saws. |
| | Car horn | Horn or klaxon of a vehicle, such as a car, van, bus, or lorry. |
| | Alarm sound | Any kind of alert signal, such as fire alarms, sirens, car alarms, and alarm clocks. |
| Human Generated | Music | Sounds originating from live bands, buskers, or recorded music played back through a speaker. |
| | Impact sound | Any transient impact sounds, and other loud impact-like sounds which may not be transient such as opening/closing of shutter, trolley wheels rolling, and dragging of chairs/tables. |
| | Shouting | Sound of a person or people shouting or screaming. |
| | Human voice | Sound of a person or people talking. Also includes coughing and sneezing. |
| Environmental | Rain | Sound due to precipitation from the sky, such as a drizzle or heavy rain. |
| | Bird | Vocalization by a bird. |
| | Ambient | Any sound not belonging to the above 10 classes. Some of the common sounds found in this class include dog barking, wind sound, squeaking baby shoes, sweeping floor, footsteps, etc |

*C. Cloud Services*

The output metadata from WASN is sent to the cloud for storage in a database and broadcast to data consumers accessing the sensor data. The database storage used is MySQL running in the cloud service provider relational database system (RDS). For relaying the sensor data to multiple data consumers, a web application running in a cloud server is used. Event metadata is recorded into our database when the edge inference detects a class having average confidence (AC) of 80% over five frames, and each window consists of one channel of 8,192 audio samples. Let $AC_x(n)$ and $n$ denote the AC of class $x$ and $n$th frame, respectively. $AC_x(n)$ is computed as

$$AC_x(n) = \frac{1}{5}\sum_{m=0}^{4} c_x(n-m), \quad (1)$$

where $c_x(n)$ is the confidence score of class $x$ at the $n$th frame. This trigger mechanism in (1) provides a more robust approach to detect a dominant sound class in the residential areas. Very short transient sound events and highly overlapping sound events are frequently encountered in residential areas. Such sounds might confuse conventional DoA estimations and lead to high localization errors. One important novelty of our algorithm is to use a sound classification model as an enabler of the DoA estimation algorithm. As such, within five frame periods, we can determine whether a dominant sound class of interest has been identified and if so, the digital signal within these frames will be located and applied to DoA estimation.

WebSocket messages are captured and stored in the database through a serverless application provided by the cloud provider. A script in the cloud service checks each WebSocket message for its formatting and content. Once the data is verified, the metadata is extracted and stored. Another serverless application is also used for creating snapshots of the stored data in the database every 10 minutes. This snapshot is formed by aggregated data grouped by hourly, daily, weekly, and monthly records. Aggregated data provides a cleaner visualization as the number of data points to plot is much reduced and a concise display of information is attained. Furthermore, using aggregated records for post-analytics accelerates visualization by avoiding loading thousands of unaggregated records directly from the database. Fig. 5 shows an example of data aggregation into an hourly record, where multiple records from a single sensor are grouped according to the date, prediction class, and hour.

### III. SOUND EVENT CLASSIFICATION AND DOA ESTIMATION

Developing a robust acoustic model for urban noise detection requires a comprehensive and representative collection of tagged training data over a period of time [38, 39, 40]. To develop an accurate DNN model for sound classification, our urban sound dataset is solely constructed from the audio recordings collected on-site by our WASN, and this dataset is used to train, validate, and test the sound classifier. Besides sound classification, the proposed system also computes A-weighted SPL, $L_{Aeq}$, azimuth, and timestamp of each detected sound class. A discussion on DoA estimation is also included in this section.

TABLE III LOCATIONS, DESCRIPTIONS, AND COMMON SOUNDS OBSERVED AT DEPLOYED LOCATIONS IN SINGAPORE

| Location | Description | Common sounds observed |
|---|---|---|
| L1 | Near market, next to a hawker centre. | Impact sounds, music playing from speakers. |
| L2 | Near playground within residential estate, near a residential carpark. | Car alarm beeping, vehicle engine sounds, aircraft sound. |
| L3 | Open-air eatery with outdoor seating, next to a carpark by road side. | Car alarm beeping, car horn, impact sounds, vehicle engine sounds, music playing from speakers. |
| L4 | Open-air pavilion located at rooftop of a building in a residential estate. | Impact sounds, music playing from speakers, aircraft sound. |
| L5 | Near basketball court, next to an open-air residential carpark. | Car alarm beeping, vehicle engine sounds, basketball bouncing. |
| L6 | Near playground within residential estate, next to a carpark. | Car alarm beeping, car horn, vehicle engine sounds, aircraft sound. |
| L7 | Near open-air eatery with outdoor seating, next to a carpark by road side. | Same as location L3. |
| L8 | Near shophouses within residential estate. | Impact sounds, music playing from speakers. Similar to location L1. |
| L9 | Within an open-air eatery, near a road. | Impact sounds, vehicle engine sounds, music playing from speakers. Similar to location L3. |

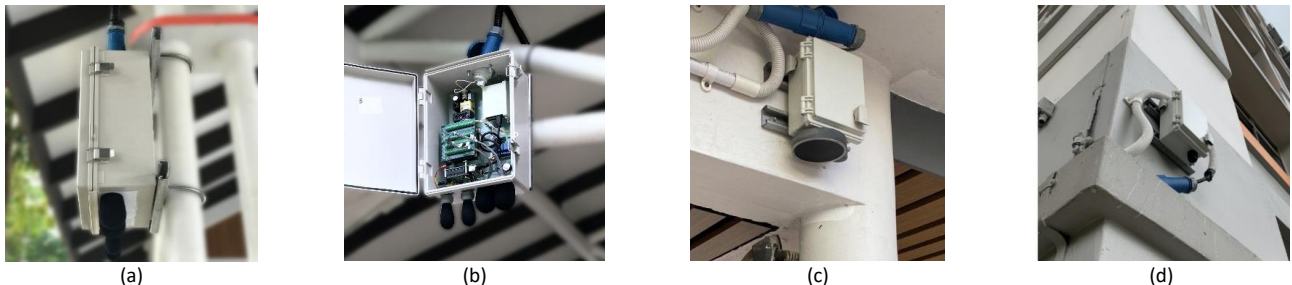

Fig. 6. WASN installed on (a) vertical pipe (single mic), (b) horizontal pipe (single mic), (c) horizontal beam (mic array), and (d) wall (single mic).

TABLE IV CLASS-WISE AUDIO LENGTH AND NUMBER OF SAMPLES IN TRAIN/VAL/TEST SPLITS

| Class | Training Set | | Validation Set | | Test Set | |
|---|---|---|---|---|---|---|
| | No. of samples | Total Audio Length (HH:MM:SS) | No. of samples | Total Audio Length (HH:MM:SS) | No. of samples | Total Audio Length (HH:MM:SS) |
| Ambient | 13,369 | 07:27:14 | 807 | 00:27:25 | 616 | 00:20:48 |
| Construction | 3,559 | 01:58:52 | 354 | 00:12:02 | 578 | 00:19:21 |
| Vehicle | 12,911 | 07:11:55 | 1,543 | 00:52:28 | 723 | 00:24:27 |
| Music | 9,008 | 05:01:05 | 343 | 00:11:39 | 651 | 00:21:51 |
| Shout | 4,311 | 02:24:30 | 490 | 00:16:44 | 511 | 00:17:16 |
| Rain | 7,670 | 04:16:18 | 357 | 00:12:13 | 610 | 00:20:28 |
| Birds | 6,985 | 03:53:14 | 433 | 00:14:53 | 390 | 00:13:11 |
| CarHorn | 1,873 | 01:02:53 | 479 | 00:16:21 | 385 | 00:12:59 |
| Alarm | 4,833 | 02:41:53 | 476 | 00:16:14 | 696 | 00:23:22 |
| ImpactSound | 4,877 | 02:43:21 | 513 | 00:17:41 | 327 | 00:11:14 |
| HumanVoice | 6,316 | 03:31:30 | 1,027 | 00:34:38 | 638 | 00:21:27 |
| **Total** | **75,712** | **42:12:45** | **6,822** | **03:52:18** | **6,125** | **03:26:24** |

### A. Taxonomy

Ten unique urban sound classes of interest are defined based on our application, and an "Ambient" class that encompasses any sound events that cannot be classified into any of the 10 classes. These sound classes can be classified into three categories, namely, machinery sound, human-generated sound, and environmental background sounds. The description of the 11 sound classes is presented in Table II.

### B. Data Collection

Our WASN are being deployed in different locations near public residential facilities. These deployed locations are chosen based on feedback from local agencies and the public.

Some common types of installations of WASN are shown in Fig. 6. In our deployments, we experienced errors in DOA estimation which are mostly due to the physical layout of the building or sheltered walkway that the WASN are installed at. Structures such as walls or metal surfaces near the microphone array introduce reflection, especially for sound events at high SPL leading to erroneous DOA estimation. Such surfaces should be avoided, or sound-absorbent material can be applied to reduce sound reflection.

Table III summarizes the deployed locations, along with a brief description of their surroundings as well as the common sounds observed at each location. It is noted that human speech and non-speech sounds and birds chirping are encountered at all locations. A total of 49 hours 31 minutes and 27 seconds of audio data are collected from nine sensor nodes deployed in nine different locations. These audio data are then annotated according to our taxonomy in Table II to provide ground-truth labels for the collected data. These annotated data form our urban sound dataset which is used is for training, validating, and testing of our model. This dataset contains 88,659 labeled sound excerpts from 11 sound classes which are drawn from our taxonomy, namely, ambient, construction, vehicle, music, shout, rain, birds, car horn, alarm, impact sound, and human voice.

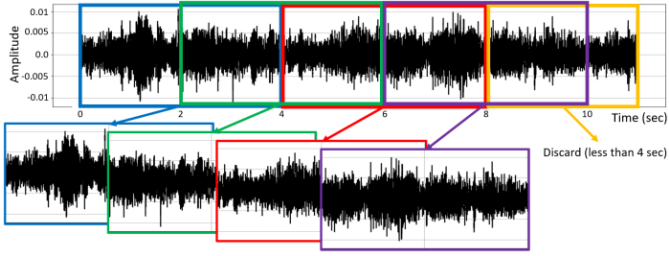

Fig. 7. Obtaining segmented audio from annotated recordings. One data sample is randomly extracted from each segment.

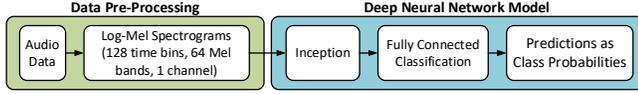

Fig. 8. Structure of proposed model architecture.

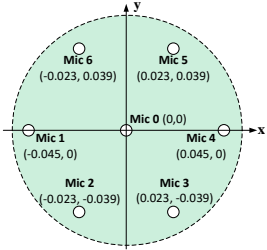

Fig. 9. Microphones on microphone array aligned to Cartesian coordinate system. Distance from mic 0 is stated in brackets and is in meters.

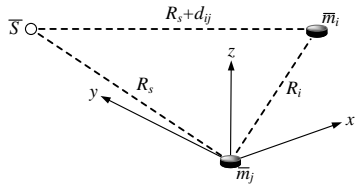

Fig. 10. Source $\bar{S}$ with respect to $i$th and $j$th microphones.

*C. Data Preprocessing and Preparation*

In this section, we discuss the pre-processing and preparation of the collected and annotated data for model training. The nine deployment sites have different distributions of the sound classes of interest. From the deployment standpoint, it would be tedious to deploy a unique sound classifier at each deployment site. To overcome this issue, we developed a scalable workflow for the collection and preparation of data in order to train a location-agnostic model. This set of data forms our urban sound database for training and evaluating a model for multi-location classification of sound. Our workflow begins with the collection of the audio recording from the sensor nodes deployed in the residential areas. After the recordings are annotated, an analysis of the distribution of the recordings at each location is performed. For our deployment, we found that the sound classes at locations L7, L8, and L9 overlapped with most of the sound classes found at locations L1 and L3. To prevent having too many similar data samples from locations L1, L3, L7, L8, and L9, only ambient data is extracted from locations L7, L8, and L9. Hence, the data samples of all sound classes in the training set are made up of recordings collected from six locations (L1-L6) and 10-minute snippets of ambient data from locations L7, L8, and L9 each.

The samples in the validation set are recordings collected from locations L1, L3, L4, L5, and L6, and the test set consists of recordings from all nine locations. Data in the training, validation, and test sets are mutually exclusive. To generate each data sample in the training, validation, and test sets, each annotated recording is divided into audio segments of length 4 seconds with an overlap of 2 seconds between segments, as shown in Fig. 7. No padding nor augmentation techniques are applied to segments containing less than 4 seconds of audio, and such segments are discarded. A 1.5 second segment is then randomly extracted from each audio segment to serve as a single data sample. Table IV summarizes the number of data samples in the training, validation, and test sets, as well as their total length. The dataset is imbalanced due to the natural difference in intra-class variation of each sound class in our taxonomy and the varying rates of these occurrences at the deployed sites.

We transform the raw audio data in each data sample into Log-Mel spectrograms as input to an InceptionNet-based network for training, validating, and testing. Log-Mel spectrograms are used because Mel filter banks are inspired by the human auditory system [41] and are commonly used for acoustic feature representation in the field of sound event classification [42, 43]. The Log-Mel spectrograms are computed from the short-time Fourier transforms (STFT) of the 1.5 second segments of raw audio with a Mel filter bank of 64 filters applied from 0 Hz to 11,025 Hz. A Hanning window of length 1,024 with 50% overlap is used in the computation of each STFT. Hence, each 1.5-second segment of single-channel raw audio in each data sample is converted to a Log-Mel spectrogram of dimension (128, 64, 1), where 128 represents the number of time bins in the STFT, 64 represents the number of filters in the Mel filter bank, and 1 represents the number of channels.

Since the 11 sound classes defined in Table IV consist of transient and stationary sounds, the audio length of each data sample should be long enough to capture at least one period of stationary sounds. At the same time, the audio length of each data sample should be short enough to maintain high resolution in time. Based on our experiments conducted with these 11 sound classes, we found that data samples at 1.5 seconds sufficiently capture the signature of the sound classes and produces good results in our validation and test sets. Based on this observation, the model input is defined to have a shape of (128, 64, 1).

As shown in Fig. 7, each audio segment is overlapped with its previous segments. Each segmented audio should be sufficiently longer than 1.5 seconds so that the randomly extracted audio samples are not duplicates of the same audio data in different segments. However, these audio segments should be kept short so that more audio samples can be extracted for training. By selecting a time interval of 4 seconds, we are able to maximize the number of critical representations from the annotated audio data, while minimizing duplicate or overlapping inputs to the classification model.

Paper ID: IoT-15711-2021    8

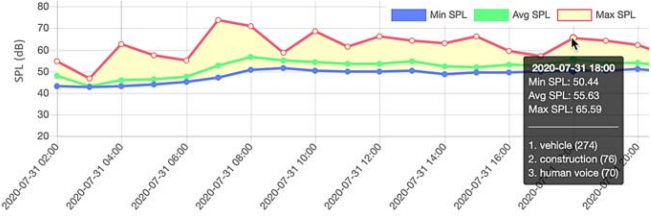

Fig. 11. SPL(minimum, average, maximum) post-analytics of location L1.

### D. Model Architecture

Model architectures such as ResNet [44], InceptionNet [45], and VGG16 [46] having millions of parameters are computationally expensive to run on our WASN, and computational cost is also incurred by other software components as shown in Fig. 1. Therefore, the model architecture should be kept minimal to reduce the required computational power, whilst providing accurate predictions. Our proposed model architecture is a shallower variant of InceptionNet [45]. As shown in Fig. 8, we combine stacks of Inception blocks and flattens the final output of the Inception blocks before feeding them into a fully-connected classification block that outputs the probabilities of a given data sample belonging to each class, as predicted by the model. The composition of each Inception block is preserved from [45], but the number of filters is reduced for the different sizes of convolutional filters in each of the Inception blocks. This approach results in a more lightweight model with a total of 335,371 parameters, which takes up 1,279 kilobytes of memory in the Raspberry Pi 3 processing unit using 32-bit floating-point representation.

### E. DoA Estimation of Dominant Sound Source

Direction-of-arrival estimation of sound sources has gained a lot of interest over the last few decades [47], as it is being used in many applications, such as teleconferencing, smart speakers, surveillance, etc. The WASN estimate the azimuth of the dominant sounds in an outdoor environment using the 7-channel microphone array, which is a concentric circular array (CCA) with a single microphone in the middle of the array, as shown in Fig. 9. Time-delay DoA (TDoA) estimation based on the generalized cross-correlation (GCC) algorithm [48] is selected for our implementation due to its computational efficiency and short decision delay. The GCC algorithm computes the time-delay of the six microphone pairs, which are formed by the six microphones at the edge of the array with the microphone in the middle of the array.

Consider that the origin of the Cartesian coordinate system is aligned to the $j$th microphone, and the time difference of a sound arriving at the $i$th and $j$th microphones is denoted as $d_{ij}$. The vectors of the $i$th microphone and source $S$ with respect to the origin are denoted as $\bar{m}_i$ and $\bar{S}$, respectively. The distance between the source and $j$th microphone is denoted as $R_s$, and the distance between the $i$th and $j$th microphones is denoted as $R_i$. An illustration of this arrangement of the microphones and sound source is shown in Fig. 10. Based on this arrangement, we have [49]:

$$\varepsilon_i = R_i^2 - d_{ij}^2 - 2R_s d_{ij} - 2\bar{S}^T \bar{m}_i, \quad \text{where } i = 2,3,\ldots N. \quad (2)$$

The term $\varepsilon_i$ accounts for any error in delay estimation with respect to the $i$th microphone. By rewriting (2) into its matrix form, we have

$$\boldsymbol{\varepsilon} = (\boldsymbol{\beta} - 2R_s \mathbf{D}) - 2\bar{S}^T \mathbf{M}, \quad (3)$$

and

$$\mathbf{M} = \begin{bmatrix} x_2 & y_2 \\ x_3 & y_3 \\ \vdots & \vdots \\ x_N & y_N \end{bmatrix}, \ \boldsymbol{\beta} = \begin{bmatrix} R_2^2 - d_{2j}^2 \\ R_3^2 - d_{3j}^2 \\ \vdots \\ R_N^2 - d_{Nj}^2 \end{bmatrix}, \ \mathbf{D} = \begin{bmatrix} d_{2j} \\ d_{3j} \\ \vdots \\ d_{Nj} \end{bmatrix}, \ \boldsymbol{\varepsilon} = \begin{bmatrix} \varepsilon_2 \\ \varepsilon_3 \\ \vdots \\ \varepsilon_N \end{bmatrix}, \quad (4)$$

where $x_i, y_i$ are the coordinates of the $i$th microphone, and $N$ denotes the number of microphones in the array. The least-squares solution of (3) minimizing $\boldsymbol{\varepsilon}$ is given by

$$\bar{S}/R_s = -(\mathbf{M}^T \mathbf{M})^{-1} \mathbf{M}^T \mathbf{D}, \quad (5)$$

assuming source $S$ is in the far-field. Equation (5) is computationally inexpensive as $(\mathbf{M}^T \mathbf{M})^{-1} \mathbf{M}^T \mathbf{D}$ can be calculated offline.

By restricting the DoA estimation to integer-valued delays $\tau$, the resolution of the DoA estimation, with respect to the broadside of the array, is

$$\theta = \sin^{-1}\left(\frac{c\tau}{f_s d}\right), \quad (6)$$

where $f_s$ is the sampling frequency of the sound classification, $c$ is the speed of sound, and $d$ is the distance between the microphones of the microphone array of the WASN. In general, the resolution of the DOA estimation is given by setting $\tau = 1$ [50] and $d = 0.045$cm [35] in (6), and the resolution of our DoA estimation is found to be 9.95°. In our implementation, we have defined the resolution of the estimated azimuth to be 10°.

## IV. VISUALIZATION AND SYSTEM PERFORMANCE

This section outlines two post-analytics methods and data visualization using SPL and detected sound classes. The urban sound classification performance of WASN that are employed in nine locations, and the mapping of detected sound classes at two locations using DoA estimation are also presented.

### A. Post-Analytics and Visualization

Fig. 11 shows an SPL visualization at location L1 on 31st July 2020. The red, blue, and green lines in Fig. 11 represent the maximum, minimum, and average SPL of the hour, respectively. For the example in Fig. 11, we observe that these three lines are trending downward after midnight, which suggests fewer human activities in the vicinity of the sensor node at location L1. On the contrary, SPL picks up from 9 am and continues to increase until 3 pm. The increased activities might be due to people visiting and having meals in the market near to the sensor node. Occurrences of high SPL are also observed between 5 pm and 7 pm, which coincides with dinner time. Fig. 12 shows the urban sound class frequency distribution at location L1 over a similar time-period as Fig. 11. From the plot, it is clear that the "Ambient" class is dominant and followed by the sound classes "HumanVoice", "Music", and "Birds". Examining the measurements between 8 am and

TABLE V PRECISION, RECALL, F1-SCORE, AND AUPRC FOR EACH OF SOUND CLASS ON VALIDATION AND TEST SETS

| Sound Class | Ambient | Constr-uction | Vehicle | Music | Shout | Rain | Birds | Car Horn | Alarm | Impact Sound | Human Voice | Averaged Accuracy | |
|---|---|---|---|---|---|---|---|---|---|---|---|---|---|
| | | | | | | | | | | | | Macro | Micro |
| *Validation Set* | | | | | | | | | | | | | |
| **Precision** | 0.930 | 0.946 | 0.951 | 0.845 | 0.923 | 0.975 | 0.937 | 0.919 | 0.966 | 0.956 | 0.907 | 0.932 | 0.934 |
| **Recall** | 0.908 | 0.986 | 0.955 | 0.968 | 0.880 | 0.972 | 0.968 | 0.929 | 0.889 | 0.895 | 0.927 | 0.934 | 0.933 |
| **F1-score** | 0.919 | 0.965 | 0.953 | 0.902 | 0.901 | 0.973 | 0.952 | 0.924 | 0.926 | 0.924 | 0.917 | 0.932 | 0.933 |
| **AUPRC** | 0.856 | 0.933 | 0.919 | 0.819 | 0.820 | 0.949 | 0.909 | 0.859 | 0.866 | 0.864 | 0.851 | 0.877 | 0.876 |
| *Test Set* | | | | | | | | | | | | | |
| **Precision** | 0.908 | 0.883 | 0.839 | 0.889 | 0.930 | 0.989 | 0.919 | 0.974 | 0.936 | 0.869 | 0.728 | 0.897 | 0.893 |
| **Recall** | 0.898 | 0.926 | 0.816 | 0.885 | 0.750 | 0.879 | 0.908 | 0.984 | 0.970 | 0.890 | 0.884 | 0.890 | 0.888 |
| **F1-score** | 0.903 | 0.904 | 0.827 | 0.887 | 0.830 | 0.931 | 0.914 | 0.979 | 0.953 | 0.879 | 0.798 | 0.891 | 0.888 |
| **AUPRC** | 0.825 | 0.824 | 0.707 | 0.799 | 0.718 | 0.881 | 0.840 | 0.960 | 0.911 | 0.779 | 0.655 | 0.809 | 0.798 |

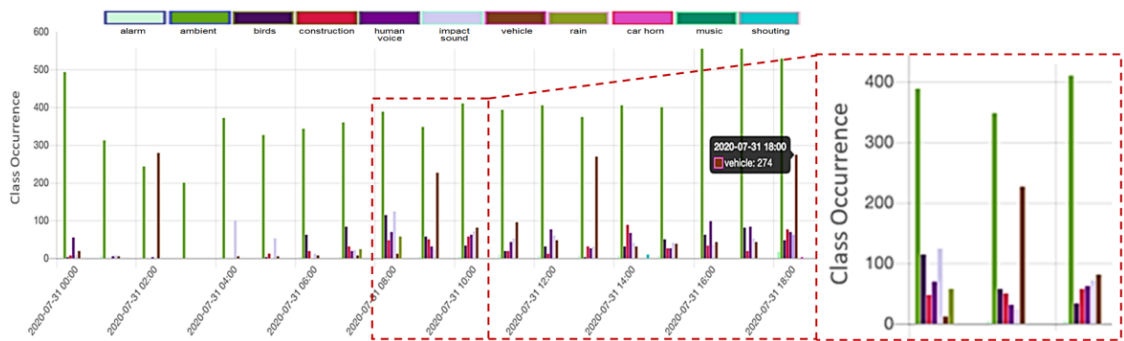

Fig. 12. Urban sound class distribution of location L1. Sequence of sound classes for each timestamp is shown at the top of the plot.

Fig. 13. Confusion matrices of trained model on (a) validation set, and (b) test set. The $(p, q)$th entry of each matrix denotes the percentage of tracks with ground-truth label $p$ that are classified by the model to be in class $q$.

10 am reveals impact and vehicle sound classes have a high occurrence at 8 am and 9 am, respectively. This observation is expected since the sensor node is installed near a market that is close to a carpark and hawker centre. Most of the detected impact sounds are unloading/loading of goods and metal wok sounds near and at the hawker centre. High occurrence of impact sound is first detected at 4 am, where hawkers were starting with food preparation and goods were delivered to the stores at the market. Between 12 am and 6 pm, high occurrences of vehicular sounds are observed at 9 am, 1 pm, and 6 pm. The high volume of vehicles at 9 am is likely due to residents leaving for work, and lunch and dinner crowds at 1 pm and 6 pm, respectively.

These SPL and sound class analytics are useful indicators on where and when the main contributing urban sound sources are affecting the environment. It should be noted that different types of post-filtering of sound event data can also be conveniently programmed over the cloud analytics platform. One such example is the filtering of prominent sound events exceeding the ambient thresholds with a noticeable change in SPL. The ability to provide another layer of data filtering in the cloud results in a more granular sieve of key sound events.

### B. Performance of Urban Sound Event Classifier

A total of 75,712 data samples is used to train the model. The model is trained over 50 epochs, with 2,366 mini-batches of size 32 per epoch. We use the Adam optimizer [51] with a learning rate of $10^{-4}$. The best model is chosen from the epoch with the highest accuracy on the validation set. The best model is evaluated on both the validation and test sets, and the confusion matrices are presented in Fig. 13. The training of the models took approximately 12 hours on four parallel GPUs using the TensorFlow framework.

The precision, recall, F1-score, and area under precision-recall curve (AUPRC) for each sound class, as well as the micro- and macro-averaged accuracies of the trained model on the validation and test sets are summarized in Table V. Given $M$ data samples, each classified into exactly one of $C$ classes, the precision $(P)$, recall $(R)$, F1-score $(F_1)$, micro-averaged accuracy ($A_{micro}$), and macro-averaged accuracy ($A_{macro}$) are, respectively, defined as

$$P = \frac{M_{TP}}{M_{TP} + M_{FP}}, \quad A_{micro} = \frac{M_{TP}}{M},$$
$$R = \frac{M_{TP}}{M_{TP} + M_{FN}}, \quad A_{macro} = \frac{1}{C}\sum_{k=1}^{C}\frac{M_{TP,k}}{M_k}, \quad (7)$$
$$F_1 = \frac{2PR}{P+R},$$

where $M_{TP}$ denotes the number of true positives, $M_{FP}$ denotes the number of false positives, $M_{FN}$ denotes the number of false negatives, $M_{TP,k}$ denotes the number of true positives belonging to class $k$, and $M_k$ denotes the number of data samples with ground truth label being class $k$. By this definition, we have $M_{TP} = \sum_{k=1}^{C} M_{TP,k}$ and $M = \sum_{k=1}^{C} M_k$. The AUPRC is calculated as a discrete estimate of the area under a curve generated by plotting the precision and recall of a model at various confidence thresholds. It is important to note that both the validation and test sets do not represent the entirety of the sounds that could potentially occur at the deployed locations of the WASN, and these sets should be thought of as a subset of all the environmental sounds at each deployed location. From Fig. 13, we observe that the model's predictions show a consistent correlation between the validation and test set results. This is especially so for the predictions made on the "Shout", "HumanVoice" and "Music" classes. For example, the model's predictions on the test set of the "Shout" class show that a significant number of samples are misclassified as "HumanVoice". Conversely, a relatively significant number of test samples of the "HumanVoice" class are also misclassified into the "Music" and "Shout" classes. Based on these observations, it is likely that there is a subtle similarity in their Log-Mel spectrogram features that the model is unable to discriminate.

*C. Mapping of Sound Events Using DoA Estimation*

While only the azimuth is estimated with the DoA algorithm, we can infer the possible regions where sound classes of interest might originate from within the location. In this sub-section, polar plots are used to illustrate detected sound classes at locations L2 and L3, and the amenities at these locations where sound classes of interest might originate from. The sound classes of interest are alarm, car horn, construction, human voice, music, shouting, and vehicle. These classes are represented by colored dots in this polar plot and are placed on seven circles around the sensor location. The sensor location is denoted by a light green circle and a white arrow pointing to 0°. The amenities at these two locations are highlighted in blue in the polar plots. We refer to this polar plot as the class locality plot. The physical installation, orientation as well as occlusion of the WASN are also discussed in this sub-section.

The sensor node at location L2 is installed on a column under a sheltered walkway as shown in Fig. 14(a), and the orientation of the microphone array is shown in Fig. 14(b). It is observed that the microphone array of the sensor node is partially blocked by the column that the sensor node is installed on. As a result, we expect very few or no sound classes to be detected at azimuths between 120° and 150°, and this conclusion is consistent with the class locality plot shown in Fig. 14(c).

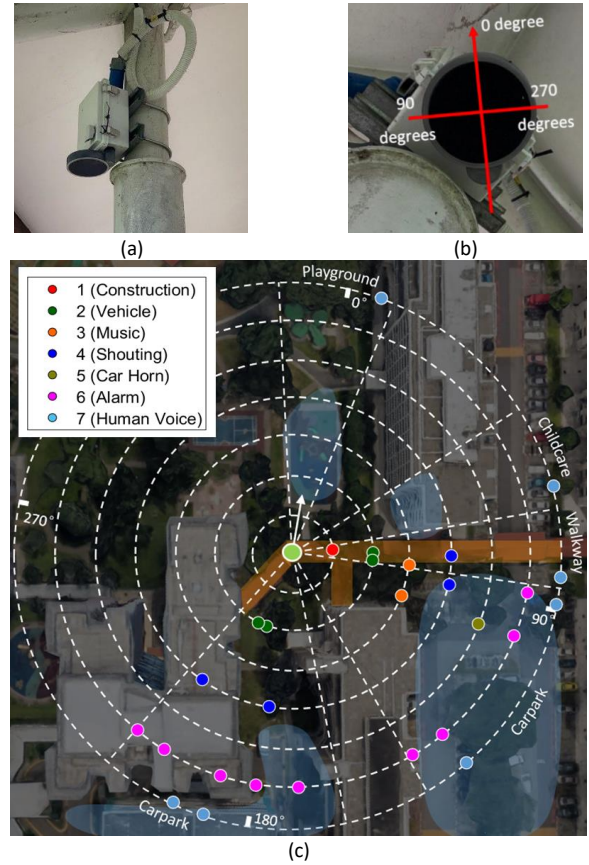

Fig. 14. Sensor node at location L2. Azimuth of 0 degrees with respect to node's orientation is indicated by arrow with green circle. (a) Node is under a sheltered walkway, (b) orientation of microphone array, and (c) locality plot of showing detected sound classes placed on concentric circles, where most inner and outer circles are numbered as 1 and 7, respectively. Concentric circles are only used for clear separation of representation of sound classes, and do not denote distance between detected sound classes and sensor node.

Human voices are detected at azimuths between 80° and 210° which are likely to be captured from passersby walking along the sheltered walkway where the sensor node is installed. Human voices are also detected at azimuths between 0° and 20°, which are in the direction of a nearby playground. Alarms (unlocking and locking of cars) from cars are detected at azimuths between 80° and 120° and between 140° and 230°, which coincide with the azimuths of two nearby carparks. In addition, car horns are detected at 110° and are generally in the direction of the carpark near the sheltered walkway. Some instances of shouting are picked up by the node at azimuths of 80°, 90°, 180°, and 210°. These instances of shouting are likely caused by individuals walking along and near the sheltered walkway. The "Music" sound class is picked up by the sensor node and sounds belonging to this class are likely to be captured from portable music players and mobile phones. Vehicular sounds are picked up at two azimuths of approximately 90° and 190°. These sounds from vehicles are likely to be originating from vehicles at the two carparks within location L2. Some construction sounds are also detected from a nearby construction site at an azimuth of 80°.

The sensor node at location L3 is positioned near an eatery beside a road, and the sensor node is installed on a column as shown in Fig. 15(a). The orientation of the microphone array is shown in Fig. 15(b), and the class locality plot of L3 is shown in Fig. 15(c). From Fig. 15(a) and 15(b), it is clear that the microphone array is blocked by the column, therefore we expect no sound events to be detected at azimuths between 180° and 270°.

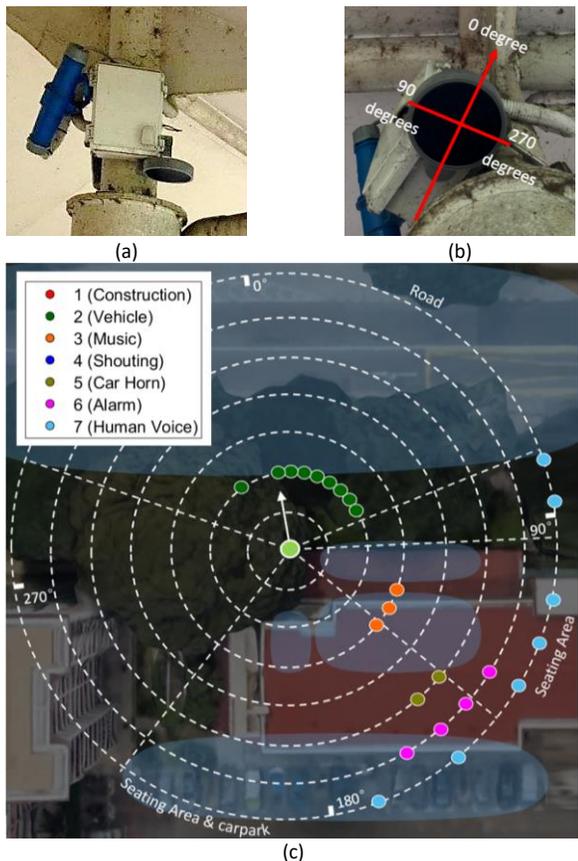

Fig. 15. Sensor node at location L3. Azimuth of 0 degrees with respect to node's orientation is indicated by arrow with green circle. (a) Node is on a column under a sheltered walkway, (b) orientation of microphone array, and (c) locality plot of showing detected sound classes placed on concentric circles, where most inner and outer circles are numbered as 1 and 7, respectively. Concentric circles are only used for clear separation of representation of sound classes, and do not denote distance between detected sound classes and sensor node.

Numerous instances of human voices are detected at azimuths between 80° to 170°, which coincide with three seating areas at the eatery near the node. There is a carpark behind the eatery (away from the road) and the alarms from cars (unlocking and locking of cars) are detected in the general direction of the carpark. Some instances of car horns are also detected in the direction of the carpark. Ringtones from mobile phones at the eating areas and sheltered walkway are detected and are correctly classified as music sounds. Many occurrences of vehicular sounds are detected from the road in front of the sensor node, at azimuths between 0° and 80° and 330° and 360°. Due to occlusion caused by the tree located at the left side of the sensor node and the column that the sensor node is installed on, no sound class of interest is detected at azimuths between 180° and 330°.

## V. CONCLUSION

Our proposed end-to-end IoT system combines machine listening at the edge with cloud services providing real-time and post descriptive analytics of urbanized residential areas. At the edge, each sensor node performs sound classification, DoA estimation, and sound measurements of the dominant sound in these areas.

We proposed a scalable workflow to collect and prepare data from nine deployed sites to train a location-agnostic model. Even though the distribution of sound classes in the deployed sites are varied, our multi-location sound classifier achieved macro-averaged accuracies of 89% and 84% in urban sound classification with the validation and test sets, respectively. For deployment of sensors at new locations, the trained sound classifier is used to reduce the time required to annotate the audio recordings from a new location. This set of new audio recordings can then be introduced to update the training set of the model, thus allowing for fine-tuning of the initially trained model.

It is increasingly challenging to manage the sensor network as the number of sensor nodes increases. Two features were implemented in the sensor nodes to assist the system administrator to manage such a sensor network. First, a self-monitoring system was implemented in each sensor node allowing the node to recover from most system failures, thus maintaining high availability of the sensor network. Second, automated remote updating of the sensor nodes was implemented to ensure that the nodes can be upgraded to the latest software for optimal performance with little or without user intervention.

Due to the ever-changing soundscape of the urban environment, there is a need for dynamic sound monitoring. Such a requirement can be fulfilled by mobile crowdsensing (MCS) [52-54]. To date, MCS has been applied in environmental and noise monitoring [55]. For future deployments, MCS using smartphones can be integrated with our sensor network to extend its coverage and to increase its spatial-temporal density, while our existing sensor network provides continuous noise monitoring at the nine deployed sites. An audience-driven based MCS system can be implemented so that residents can request ad-hoc noise monitoring at new locations, and this functionality is particularly useful to monitor noise levels of celebrations during festive seasons and at temporary construction sites.

The current sensor network is not designed specifically for low power consumption and has an average power consumption of approximately 6W. Further refinements to reduce the power consumption of the WASN are required so that a denser sensor network can be deployed. Some possibilities to reduce power consumption include deploying a solar-powered system for the WASN and adopting the long-range wide-area-network (LoRaWAN) to enable low-power and long-range communication among WASN. The LoRa gateway can also be used to reduce data rate and cost of communication among WASN. In addition, there is a need to vary the granularity of the classification types in different locations to better capture the likely dominant sound events in a particular location.

Various extensions to this system are currently underway to further enhance its capability. One such extension is to include a multi-label sound event localization and detection (SELD) system [56] that has been reported in the recent DCASE 2020 competition.

Our proposed system also offers customizable data visualization to monitor any sensor node in real-time and delivers post descriptive analytics using aggregated metadata of the sound received from our sensor network. As the proposed edge-cloud system is seamlessly integrated with features for over-the-air (OTA) upgrading and self-monitoring functionalities to increase system robustness, and scalable to deploy at new locations, there are many opportunities to further integrate other types of environmental sensors, which complement and combine with each other into a more holistic sensor network to serve different application needs.

ACKNOWLEDGMENT

This research/project is supported by the National Research Foundation and the Smart Nation Digital Government Office, Prime Minister's Office, Singapore under the Translational R&D for Smart Nation (TRANS Grant) Funding Initiative. The research work on direction of arrival estimation is also supported by the Singapore Ministry of Education Academic Research Fund Tier-2, under research grant MOE2017-T2-2-060.

The authors would like to thank Santi Peksi for her invaluable support in collecting and processing audio recordings for our urban sound dataset.

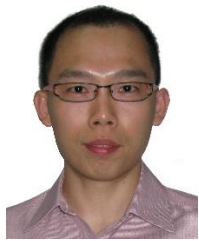

**Ee-Leng Tan** received his B.Eng. (1st Class Hons) and Ph.D. degrees, both in Electrical and Electronic Engineering from the Nanyang Technological University (NTU) in 2003 and 2012, respectively. Currently, he is with NTU as a Senior Research Fellow. His research interests include real-time signal processing, perceptual image processing, 3D and directional audio, and machine learning in medical imaging, and his work has translated into 3 patents.

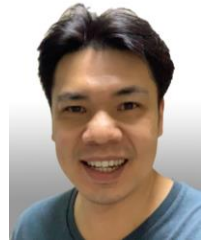

**Furi Andi Karnapi** is currently a senior research engineer under the Smart Nation Translational Lab at Nanyang Technological University (NTU) in Singapore. He received the M.Eng. in Information Engineering from the School of Electrical and Electronic Engineering, NTU, Singapore in 2003. Prior to joining NTU, he had experience in technical sales/product marketing and application engineering in semiconductor and electronics manufacturing. His current research interests are IoT and embedded design, cloud-based applications, audio processing, DSP, and machine learning in embedded processors.

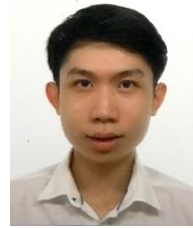

**Linus Ng** received a B.Eng. degree in Electrical and Electronic Engineering from Nanyang Technological University (NTU), Singapore, in 2017. He is currently pursuing a M.Eng degree in Electrical Engineering at NTU and is also a Research Engineer under the Smart Nation Translational Lab in NTU. His research interests include deep learning for acoustic scenes, environmental sound classification, and sound event detection.

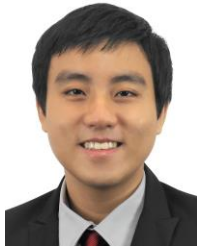

**Kenneth Ooi** received a B.Sc. degree in Mathematical Sciences from Nanyang Technological University (NTU), Singapore, in 2019. He is currently pursuing a Ph.D. degree in Electrical Engineering at NTU, Singapore. His research interests include deep learning for acoustic scene and event classification, as well as applications of machine learning to the field of soundscape research.

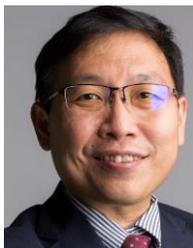

**Woon-Seng Gan** (Senior Member, IEEE) received his B.Eng. (1st Class Hons) and Ph.D. degrees, both in Electrical and Electronic Engineering from the University of Strathclyde, UK in 1989 and 1993 respectively. He is currently a Professor of Audio Engineering in the School of Electrical and Electronic Engineering (EEE) in Nanyang Technological University (NTU). He also served as the Head of the Information Engineering Division in the School of EEE at NTU from 2011-2014. He is currently the Director of the Smart Nation Lab at NTU.

He is a Fellow of the Audio Engineering Society (AES), a Fellow of the Institute of Engineering and Technology (IET), and a Senior Member of the IEEE. He also served as an Associate Editor of the IEEE/ACM Transaction on Audio, Speech, and Language Processing (TASLP; 2012-15); and Senior Area Editor of the IEEE Signal Processing Letters (SPL; 2016-2019).